%% file: leschke-update.tex
\documentclass{gen-p-l}
\usepackage{mathsym2}
\newtheorem{theorem}{Theorem}[section]
\newtheorem{lemma}[theorem]{Lemma}
\newtheorem{proposition}[theorem]{Proposition}
\theoremstyle{definition}
\newtheorem{definition}[theorem]{Definition}

\newtheorem{examples}[theorem]{Examples}
\theoremstyle{remark}
\newtheorem{remark}[theorem]{Remark}
\newtheorem{remarks}[theorem]{Remarks}
\newtheorem{assumption}[theorem]{Assumption}
\numberwithin{equation}{section}
\newcounter{numcount}
\makeatletter
\newenvironment{nummer}%
    {\let\curlabelspeicher\@currentlabel%
     \newcommand{\labelnummer}{\mbox{(\roman{numcount})}}%
     \begin{list}{\labelnummer}{\usecounter{numcount}\leftmargin0pt%
                  \topsep1ex\partopsep2ex\parsep0pt\itemsep1ex%
                  \labelwidth3.5em\itemindent2.5em\labelsep1em}%
     \let\saveitem\item%
     \def\item{\saveitem%
               \def\@currentlabel{\curlabelspeicher\hspace{.2em}\labelnummer}%
               \let\label\bemlabel}}%
   {\end{list}}%
\def\@cite#1#2{{%
  \m@th\upshape\mdseries[{#1\if@tempswa, #2\fi}]}}
\makeatother
\makeatletter
\RequirePackage{amsthm}[1996/09/24]
\def\@swapped#1#2{#2%
  \@ifnotempty{#1}{\@addpunct{.}\quad#1\unskip}}
\def\thmhead@plain#1#2#3{%
  \thmname{#1}\thmnumber{\@ifnotempty{#1}{ }\@upn{#2}}%
  \thmnote{ \textmd{\upshape(#3)}}}
\def\swappedhead@plain#1#2#3{%
  \thmnumber{\@upn{#2}}\thmname{\@ifnotempty{#2}{. }#1}%
  \thmnote{ \textmd{\upshape(#3)}}}
\def\th@plain{%
  \let\thm@indent\indent
  \thm@headfont{\bfseries}
  \let\thmhead\thmhead@plain \let\swappedhead\swappedhead@plain
  \thm@preskip.5\baselineskip\@plus.2\baselineskip
                                    \@minus.2\baselineskip
  \thm@postskip\thm@preskip
  \itshape
}
\def\th@definition{%
  \let\thm@indent\indent
  \thm@headfont{\bfseries}
  \let\thmhead\thmhead@plain \let\swappedhead\swappedhead@plain
  \thm@preskip.5\baselineskip\@plus.2\baselineskip
                                    \@minus.2\baselineskip
  \thm@postskip\thm@preskip
  \upshape
}
\def\th@remark{%
  \let\thm@indent\indent
  \thm@headfont{\bfseries}
  \let\thmhead\thmhead@plain \let\swappedhead\swappedhead@plain
  \thm@preskip.5\baselineskip\@plus.2\baselineskip
                                    \@minus.2\baselineskip
  \thm@postskip\thm@preskip
  \upshape
}
\makeatother
\newfont{\fiverm}{cmr5 at 5pt} 

\catcode`@=11 \catcode`!=11

\expandafter\ifx\csname fiverm\endcsname\relax
  \let\fiverm\fivrm
\fi
  
\let\!latexendpicture=\endpicture 
\let\!latexframe=\frame
\let\!latexlinethickness=\linethickness
\let\!latexmultiput=\multiput
\let\!latexput=\put
 
\def\@picture(#1,#2)(#3,#4){%
  \@picht #2\unitlength
  \setbox\@picbox\hbox to #1\unitlength\bgroup 
  \let\endpicture=\!latexendpicture
  \let\frame=\!latexframe
  \let\linethickness=\!latexlinethickness
  \let\multiput=\!latexmultiput
  \let\put=\!latexput
  \hskip -#3\unitlength \lower #4\unitlength \hbox\bgroup}

\catcode`@=12 \catcode`!=12
\input{pictex}

\catcode`@=11 \catcode`!=11
  
\let\!pictexendpicture=\endpicture 
\let\!pictexframe=\frame
\let\!pictexlinethickness=\linethickness
\let\!pictexmultiput=\multiput
\let\!pictexput=\put

\def\beginpicture{%
  \setbox\!picbox=\hbox\bgroup%
  \let\endpicture=\!pictexendpicture
  \let\frame=\!pictexframe
  \let\linethickness=\!pictexlinethickness
  \let\multiput=\!pictexmultiput
  \let\put=\!pictexput
  \let\input=\@@input   
  \!xleft=\maxdimen  
  \!xright=-\maxdimen
  \!ybot=\maxdimen
  \!ytop=-\maxdimen}

\let\frame=\!latexframe

\let\pictexframe=\!pictexframe

\let\linethickness=\!latexlinethickness
\let\pictexlinethickness=\!pictexlinethickness

\let\\=\@normalcr
\catcode`@=12 \catcode`!=12

\DeclareMathOperator{\RL}{L}
\DeclareMathOperator{\Ri}{i}

  \def\esssup{\mathop{\hbox{ess$\,$sup}}}
  
\makeatletter
\def\norm{\@ifnextchar[{\normmit}{\normohne}}
\def\normohne#1{\left\lVert#1\right\rVert}
\def\normmit[#1]#2{\left\lVert #2 \right\rVert_{#1}}
\makeatother
\makeatletter

\def\intr{\@ifnextchar[\@@intr\@intr}
\def\@@intr[#1]#2{\int_{\rz^{#1}}\!\d^{#1} #2\;}
\def\@intr#1{\int_{\rz}\!\d #1\;}

\def\et{ and }
\def\bti#1{\emph{#1}}
\def\ti#1{\textit{#1},}
\def\z#1#2#3#4{#1 \textbf{#2} (#4) #3.}
\def\au#1#2{#1 #2}

\begin{document}
\title[Multiformity of magnetic Lifshits tails]{The 
  multiformity of Lifshits tails
 caused by\\ random Landau Hamiltonians
 with repulsive impurity potentials of different decay at infinity}
\author{Thomas Hupfer}
\address{Institut f{\"u}r Theoretische Physik,
  Universit{\"a}t  Erlangen-N{\"u}rnberg,
  Staudtstra{\ss}e~7,
  D-91058 Erlangen, Germany.
  }
\email{thomas.hupfer@theorie1.physik.uni-erlangen.de}
%
\author{Hajo Leschke}
\email{hajo.leschke@theorie1.physik.uni-erlangen.de}
\author{Simone Warzel}
\email{simone.warzel@theorie1.physik.uni-erlangen.de}
%
\subjclass{82B44, 82D30}
\date{\today.}
%
%
\begin{abstract}
For a charged quantum particle in the Euclidean plane subject to a 
perpendicular constant magnetic field and repulsive impurities, randomly
distributed according to Poisson's law, we determine the leading low-energy fall-off of the 
integrated density of states in case the single-impurity potential has either
super-Gaussian or regular sub-Gaussian long-distance decay.
The forms of the resulting so-called magnetic Lifshits tails reflect the great variety of these
decays.
On the whole, we summarize, unify, and generalize results in previous works of
K.\ Broderix, L.\ Erd\H{o}s, D.\ Hundertmark, W.\ Kirsch, and ourselves.
\end{abstract}
\maketitle
\vspace*{-0.1cm}
\section{Introduction}
The energy spectrum of a spinless quantum particle with mass $m>0$ and electric charge $Q\neq0$
in the
Euclidean plane $\rz^2$ subject to a perpendicular constant magnetic
field of strength $B>0$
is explicitly known since the early works of Fock \cite{Foc28}
and Landau \cite{Lan30}.
It consists only of isolated harmonic-oscillator like eigenvalues $\varepsilon_0$, $3\varepsilon_0$,
$5\varepsilon_0$, $\dots$ of infinite degeneracy, where $\varepsilon_0:=\hbar\, |Q| B /2m$ is the so-called {\it lowest Landau level}
and $2\pi\hbar>0$ is Planck's constant. 
The underlying ``magnetic'' Schr\"odinger operator
\begin{equation}
  H(0) \,:=\, \frac{1}{2m} \left[ %
                         \left(\Ri \hbar \frac{\partial}{\partial x_1} -%
                             \frac{Q B}{2}x_2\right)^2 +%
                         \left(\Ri \hbar \frac{\partial}{\partial x_2} +%
                             \frac{Q B}{2}x_1\right)^2 %
                       \right], %
\end{equation}
acting on the Hilbert-space  $\RL^{2}(\rz^{2})$
of Lebesgue square-integrable, complex-valued
functions on $\rz^2$, is often referred to as the \emph{Landau Hamiltonian}.
Here $\Ri=\sqrt{-1}$ stands for the imaginary unit and $(x_1,x_2)$ for
the pair of Cartesian co-ordinates of a given point
$x \in \rz^2$ interpreted as the classical position of the
particle. 
The spectral resolution of $H(0)$ may be written as 
\begin{equation}
 H(0) = \varepsilon_0 \sum^{\infty}_{n=0} (2n+1) P_n.
\end{equation}
The orthogonal projection $P_n$ associated with the \emph{$n$th Landau 
level} $(2n+1)\varepsilon_0$ is an integral operator with a 
continuous kernel. Its diagonal
$P_n(x,x)=1/(2\pi\ell^2)$, given in terms of the so-called \emph{magnetic length}
$\ell:=\sqrt{\hbar/|Q|B}$, is naturally interpreted as
the degeneracy of the $n$th Landau level
per area. 

In recent decades, the fabrication of low-dimensional semiconductor 
micro-structures and micro-devices as well
as the discovery of the (integer) quantum Hall effect have stimulated 
investigations of the so-called \emph{random Landau Hamiltonian}
\begin{equation} \label{schoperator}
  H(V_\omega) := H(0) + V_\omega ,
\end{equation}
where $V_\omega$ is some
random potential modelling the interaction of the particle with irregularly distributed
impurities. An important issue is to understand the spectral properties of 
the perturbed operator (\ref{schoperator}).

In this paper we choose $V_\omega$ to be a repulsive 
Poissonian potential
\begin{equation} \label{Poisson}
  V_\omega(x) := \sum_j U\left(x-q(\omega,j)\right), \qquad U \geq 0.
\end{equation}
Here for a given realization $\omega \in \Omega$ of the randomness the
point $q(\omega,j)\in \rz^2$ stands for the position of the $j$th impurity 
repelling the particle at $x\in \rz^2$ by a positive potential $U$ which neither
depends on $\omega$ nor on $j$. We assume that the single-impurity potential
$U$ is strictly positive on some
non-empty open set in $\rz^2$. The impurities are supposed to be distributed 
``completely at random'' on the plane. More precisely, 
the probability of
simultaneously finding $M_1, M_2, \dots , M_K$ impurity points
in respective pairwise disjoint (Borel) subsets
$\Lambda_1, \Lambda_2, \dots , \Lambda_K \subset \rz^2$
is given by the product
$\prod_{k=1}^K \e^{-\varrho \left| \Lambda_k \right|}
\left(\varrho \left| \Lambda_k \right| \right)^{M_k}/M_k!\,$, where
$ \left|\Lambda_k\right|:=\int_{\Lambda_k}\! \d^2 x$ is the area of $\Lambda_k$ and
the parameter $\varrho >0$ is the mean concentration of impurities.

The simplest but physically important
spectral characteristics of the random Landau Hamiltonian (\ref{schoperator})
is its \emph{integrated density
of states} $N: E \mapsto N(E)$. 
Roughly speaking, $N(E)$ is the averaged number of 
energy levels per area below a given energy $E\in \rz$.
Under rather weak assumptions on $U$ it can be shown that 
both the spectrum of $H(V_\omega)$ and the set of growth points of $N$
coincide for almost all $\omega\in\Omega$ with the half-line $[\varepsilon_0,\infty[$.

\begin{figure}[tb]\label{bild}
\begin{center}
\parbox[t]{12cm}{%
\input{test2.pictex.tex}}
\caption{\footnotesize
Qualitative plots of the integrated density of states $N(E)$ 
of the random Landau Hamiltonian as a function of the energy $E$.
The dashed line corresponds to the case without impurities ($\varrho =0$).
The upper and lower solid line correspond to the presence of repulsive impurities
with mean concentration $\varrho_1>0$ and $\varrho_2 > \varrho_1$, 
respectively. By definition, the Lifshits tail is the leading 
asymptotic behaviour of $N(E)$ as $E$ approaches 
the lowest Landau level $\varepsilon_0$ from above.}
\end{center}
\end{figure}
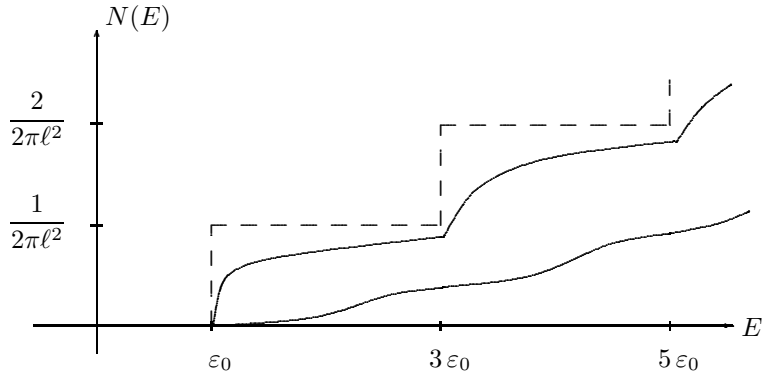

For the unperturbed Landau Hamiltonian $H(0)$ the graph of $N$ looks like a
staircase, where the distance of successive steps as well as their height are
proportional to $B$, confer Figure~\ref{bild}.
The presence of repelling impurities will lift the degeneracy of the Landau levels and
reduce the values 
of $N$. Moreover, since the impurities are randomly distributed, the steps are
expected to be smeared out the more the stronger the randomness is.
For a sketch of the resulting graph of $N$ for a given $U$ and 
two different values of $\varrho$ see again Figure~\ref{bild}.
But the reader should notice that nobody really knows what $N$ is looking like
for $\varrho>0$.
In particular, there is not even a proof that $N$ will become continuous for sufficiently
strong randomness.
  
To our knowledge, 
the only rigorous results available so far for $N$ of the random
Landau Hamiltonian (\ref{schoperator}) with a Poissonian potential (\ref{Poisson})
concern its asymptotic high-energy 
growth \cite{Mat93,Uek94} and low-energy fall-off \cite{BHKL95,Erd98,HLW99}.
The first asymptotics is neither affected by the impurities nor the magnetic field
and given by 
\begin{equation}
N(E) \, \sim \, \frac{m}{2\pi\hbar^2}\, E  = 
\frac{(1/2\pi \ell^2)}{2 \varepsilon_0} \, E \qquad \quad (E \to \infty),
\end{equation}
which is consistent with a famous result of Weyl \cite{Wey12}.
The asymptotic behaviour of $N$
near the bottom of the almost-sure spectrum $[\varepsilon_0,\infty[$, that is, the 
behaviour of $N(\varepsilon_0+E)$
for $E\downarrow 0$ is more complicated and harder to obtain. 
Since the presence of the impurities should rarefy especially
low-lying energy levels, one expects that the values of $N$ 
near $\varepsilon_0$ are 
dramatically diminished such that $N$ drops down to zero continuously. 
The resulting leading low-energy fall-off is
commonly referred to as a \emph{Lifshits tail} in honour of the 
theoretical physicist I.\ M.\ Lifshits, who was the first to 
develop a quantitative theory in the case $B=0$ \cite{Lif64,LGP88}.
The form of the Lifshits tail mainly depends on the decay of the
impurity potential $U(x)$ for long (Euclidean) distance $|x|:=\sqrt{x_1^2 +x_2^2}$. 
This stems from the fact that low-lying energy levels mainly originate
in large regions in $\rz^2$ without impurities. As a consequence, there the 
particle's potential energy is solely due to impurities 
outside these regions
and hence to the
long-distance tail of $U$.
 
The purpose of the present contribution is to 
summarize, unify, and generalize previous results in \cite{BHKL95,Erd98,HLW99}.
Thereby we illustrate that in the presence of a magnetic field 
different 
long-distance decays of the 
single-impurity potential lead to a
huge multiformity of Lifshits tails.  
Loosely speaking, there are more Landau-Lifshits tails than Lifshits tails.

The paper is organized as follows. The next section provides conditions for $U$ to enable precise
definitions of the random Landau Hamiltonian with a Poissonian potential and
of its integrated density of states. The main results on Lifshits tails in 
magnetic fields are contained in two theorems presented in Section~\ref{Lifshits}.
The first theorem quotes the result for so-called super-Gaussian decay to be found in \cite{Erd98,HLW99}.
The second theorem identifies the Lifshits tails for all impurity potentials with 
so-called regular sub-Gaussian decay.
The proof of the latter theorem is given in Section~\ref{Proof}.
Open problems related to sub-Gaussian but not regular decay are briefly 
discussed at the end of Section~\ref{Lifshits}.
For convenience, Appendix~\ref{AppendixRegVar} compiles useful facts about regularly varying 
functions which underly our definition of regular decay. For completeness, 
Appendix~\ref{AppendixTauber} presents a
Tauberian theorem needed in Section~\ref{Proof}.
\section{Basic assumptions and definitions}
Throughout this paper we require the single-impurity potential $U$ to satisfy
\begin{assumption}\label{Genass}
$U$ is positive, integrable, and locally
square integrable 
\begin{equation} \label{general}
U \geq 0, \qquad\quad U \in {\rm L}^1(\rz^2) \cap {\rm L}^2_{\rm loc}(\rz^2),
\end{equation}
and strictly positive on some non-empty open set in $\rz^2$.
\end{assumption}
\begin{remarks}
\begin{nummer}
\item
Assumption~\ref{Genass} guarantees that the Poissonian
potential (\ref{Poisson}) can be rigorously defined as a positive, measurable,
ergodic random field on $\rz^2$ with an underlying complete probability space 
$\left(\Omega,{\mathcal A},{\mathbb P}\right)$.
\item
By Assumption~\ref{Genass} and \cite[Theorem 1.15]{CFKS87}
the random Landau Hamiltonian 
$H(V_\omega)$, informally given by (\ref{schoperator}), is for ${\mathbb P}$-almost all $\omega \in \Omega$
defined as a  self-adjoint ``magnetic'' Schr\"odinger operator on ${\rm L}^2(\rz^2)$. 
Moreover, $\omega \mapsto H(V_\omega)$ is 
measurable and ergodic with respect to (magnetic) translations. 
\end{nummer}
\end{remarks}
The object of interest in this paper, the integrated density of states $N$,
may be defined by the expectation value
\begin{equation} \label{IDOS}
N(E) := \int_\Omega \! \d {\mathbb P}(\omega)\, \Theta\left(E-H(V_\omega)\right)(x,x),
\qquad E \in \rz.
\end{equation}
Here the mapping
$\rz^2 \times \rz^2 \ni (x,y) \mapsto \Theta\left(E-H(V_\omega)\right)(x,y)$,
denotes the continuous
integral kernel of the spectral projection $\Theta\left(E-H(V_\omega)\right)$
associated with the half-line $]-\infty,E[$.
In fact, $N$ is the distribution function of a positive Borel
measure on the real line $\rz$ with topological
support equal to $[\varepsilon_0, \infty[$,
the spectrum of $H(V_\omega)$ for ${\mathbb P}$-almost all $\omega$.
\begin{remarks}
\begin{nummer}
\item
Theorem 6.1 and Remark 6.2 (ii) in \cite{BHL99} imply that spectral projections 
of $H(V_\omega)$ indeed possess continuous integral kernels, see also \cite[Lemma 3.1]{Uek94}.
\item
Due to (magnetic) translation invariance the right-hand side of (\ref{IDOS})
is independent of $x \in \rz^2$.
\item
Definition (\ref{IDOS}) coincides with the more physical one by means of a spatial average
in the macroscopic limit \cite{Mat93,Uek94}. More precisely, by restricting $H(V_\omega)$ to
a bounded open square in $\rz^2$ with zero Dirichlet boundary conditions,
one defines its finite-area integrated density
of states to be the number of eigenvalues below $E$ divided by the area 
of the square.
Thanks to ergodicity of $V_\omega$, this quantity becomes non-random
in the macroscopic limit of an unbounded square 
and coincides for ${\mathbb P}$-almost all $\omega \in \Omega$
with $N(E)$ except for the at most countably many discontinuity points of $N$.
\end{nummer}
\end{remarks}
For a general background concerning the mathematical theory of random Schr\"o\-ding\-er operators,
see \cite{Kir89,CaLa90,PaFi92}.
\section{Lifshits tails in magnetic fields}\label{Lifshits}
As was already pointed out in the Introduction, 
the form of the Lifshits tail depends on the long-distance decay of the
impurity potential $U$. 
We may classify the different decays of impurity potentials 
by defining the following three pairwise disjoint classes.
\begin{definition}
Let $U$ be a positive measurable function on $\rz^2$ decaying to zero at infinity,
$\inf_{R>0} \, \esssup_{|x|>R} U(x)=0$. Then $U$ is said
to have \emph{super-Gaussian decay} if
\begin{equation}\label{SuperGauss}
   \inf_{R>0} \, \esssup_{|x|>R}\,\frac{\log U(x)}{|x|^2} = - \infty,
\end{equation}
\emph{Gaussian decay} 
if there exists a constant $\lambda \in]0,\infty[$ such that
\begin{equation}\label{Gauss}
   \inf_{R>0} \, \esssup_{|x|>R}\,\frac{\log U(x)}{|x|^2} = - \frac{1}{\lambda^2},
\end{equation}
and \emph{sub-Gaussian decay} if
\begin{equation}\label{SubGauss}
   \inf_{R>0} \, \esssup_{|x|>R}\,\frac{\log U(x)}{|x|^2}  = 0.
\end{equation}
\end{definition}
\begin{remarks}
\begin{nummer}
\item
The condition $U\in {\rm L}^1(\rz^2)$ is not sufficient to guarantee
$\inf_{R>0}$\hspace{2pt}$\esssup_{|x|>R} U(x)=0$. 
\item
Roughly speaking, $U$ has super-Gaussian or sub-Gaussian decay if
its upper envelope decays faster or slower than any Gaussian. 
The borderline case of Gaussian decay occurs when the 
upper envelope of $U$
decays like a Gaussian with some definite decay length $\lambda$.
Adopting the convention $\log 0 := - \infty$ the class of impurity potentials with
super-Gaussian decay also includes all compactly
supported $U$. 
In case the long-distance decay of $U$ is not isotropic, the above definitions 
always detect the slowest decay in whatever direction it may occur.
\item
Clearly, for continuous $U$ one may replace in the above formulae 
the Lebesgue-essential upper limit
$\inf_{R>0} \, \esssup_{|x|>R}$ simply by
$\limsup_{|x| \to \infty}$.
\end{nummer}
\end{remarks}
\subsection{Lifshits tails caused by super-Gaussian and Gaussian decay}
Interestingly enough, all impurity potentials $U$ with super-Gaussian
decay cause the same Lifshits tail. More precisely, one has
\begin{theorem}\label{ThmQuantum}
Let the positive impurity potential $U$ be in ${\rm L}^2_{\rm loc}(\rz^2)$, be
strictly positive on some non-empty open set in $\rz^2$,
and have super-Gaussian decay (\ref{SuperGauss}). Then the integrated density of
states has power-law fall-off to zero at $\varepsilon_0>0$ in the sense that
\begin{equation}\label{QLT}
\log N(\varepsilon_0+E) \, \sim \,  \log\big( E^{2 \pi \varrho  \ell^2}\big) 
\, \sim \, - 2\pi \varrho \, \ell^2 \left|\log E\right| 
\qquad \quad (E \downarrow 0).
\end{equation}
\end{theorem}
\begin{remarks}
\begin{nummer}
\item
Here and in the following we write $F(E) \sim G(E)$ ($E \downarrow 0$) and 
$F(t) \sim G(t)$ ($t \to \infty$) 
as a short-hand
for \emph{asymptotic equivalence} (at the origin and at infinity, respectively) 
of two real-valued functions $F$ and $G$ in the sense that 
$\lim_{E\downarrow 0} F(E)/G(E) =1$ and
$\lim_{t \to \infty} F(t)/G(t) =1$, respectively.
\item
The Lifshits tail (\ref{QLT}) exhibits a genuine quantum character because it depends, through 
the magnetic length $\ell$, on Planck's constant and the magnetic field. 
\item
The exponent $2 \pi \varrho \, \ell^2$ in (\ref{QLT}) 
is just the mean number of 
impurities in a disk of radius $\sqrt{2} \ell$. Depending on whether this
number is smaller or larger than one, $N$ exhibits a root-like or 
true power-law fall-off (in the logarithmic sense). 
Both cases are sketched in Figure~\ref{bild}.  
\item
The theorem is proven in \cite{HLW99} where we heavily rely on L. Erd\H{o}s'
result \cite{Erd98} for compactly supported $U$. For its involved proof he
developed a version of the ``method of enlargement of obstacles'' \cite{Szn98}.
\end{nummer}
\end{remarks}
In case of Gaussian decay (\ref{Gauss}) the question as to how the leading low-energy fall-off
of $N$ looks like is open. We only have the following result
\begin{equation}\label{NGauss}
 -\pi \varrho (\lambda^2 + 2 \ell^2) \leq 
  \liminf_{E \downarrow 0} \frac{\log N(\varepsilon_0 + E)}{\left|\log E \right|}
  \leq \limsup_{E \downarrow 0} \frac{\log N(\varepsilon_0 + E)}{\left|\log E \right|} \leq
  -2 \pi \varrho \, \ell^2.
\end{equation}
This follows from estimating the long-distance decay of $U$ from below and above
by some super-Gaussian decay and by the definite Gaussian decay
$\lim_{|x|\to \infty} |x|^2/$\hspace{2pt}$\log U(x) = -\lambda^2$, respectively. 
If $U$ already has this definite Gaussian decay, the upper bound $-2 \pi \varrho \,\ell^2$
in (\ref{NGauss}) may be sharpened to  $-\pi \varrho \max\{\lambda^2, 2\ell^2\}$.
Confer \cite{HLW99}.
\subsection{Lifshits tails caused by sub-Gaussian decay}
In contrast to super-Gaussian decay, sub-Gaussian decay of the
impurity potential $U$ allows for a great variety of Lifshits
tails whose details sensitively depend on the details of the decay of $U$.
A wide class of impurity potentials able to illustrate the multiformity of 
Lifshits tails caused by different sub-Gaussian decay is contained in the 
class of functions having a definite isotropic decay in the sense of  
\begin{definition}\label{AssRegVar}
A positive measurable function $U$ on $\rz^2$ is said to have 
\emph{regular decay}, or more specifically, a
\emph{regular $(F,\alpha)$-decay} if 
\begin{equation}\label{AbfallU}
\lim_{|x| \to \infty}\, \frac{F(1/U(x))}{|x|} = 1, 
\end{equation}
for some positive function $F$, 
which is regularly varying of index
$1/\alpha \in [0,\infty[$, in symbols $F \in R_{1/\alpha}$, and strictly   
increasing towards infinity, $F(t) \to \infty$ as $t \to \infty$.
Here we adopt the conventions $1/\infty~:=0$ and $1/0~:=\infty$.
\end{definition} 
\begin{remarks}\label{nummer1}
\begin{nummer}
\item
For the definition and some useful properties of regularly varying functions,
see Appendix~\ref{AppendixRegVar}. 
\item
Since $F$ is strictly increasing its inverse $F^{-1}$ exists as a positive function
on the half-line $[\inf_{t >0} F(t), \infty[$ and is also strictly increasing.
Equation (\ref{AbfallU}) therefore requires $U$ to be asymptotically
of the form $U(x)=1/F^{-1}(|x|(1+o(1)))$ as $|x| \to \infty$, where 
``little oh'' $o(1)$ stands for any function decaying to zero.
Moreover, since $F \in R_{1/\alpha}$ it follows that $F^{-1} \in R_\alpha$
(confer Appendix~\ref{AppendixRegVar}).
In case $\alpha < \infty$
one therefore shows with the help of (\ref{eq:repRegVar}) that  
every $U$ with regular decay is asymptotically of the form 
$U(x)= |x|^{-\alpha}f(|x|)(1+o(1))$ 
as $|x| \to \infty$, with some $f \in R_0$.
\item
Although $F$ is required to be strictly increasing, $U$ does not 
necessarily decay 
monotonously. However, the leading long-distance decays of its lower and upper 
envelope have to coincide.
For example, $U(x)=g_0 \, |x|^{-\alpha} \left[2+ (\lambda/|x|)\sin(|x|/\lambda)\right]$
with some constants $g_0$, $\lambda$, $\alpha>0$ has regular decay, but
$U(x)=g_0 \, |x|^{-\alpha} \left[2+\sin(|x|/\lambda)\right]$ 
has not.
\item
For a given $U$ with regular decay the function $F$ in Definition~\ref{AssRegVar} is 
determined only up to asymptotic equivalence at infinity. This freedom may be used to 
choose $F$ smooth. 
\item
A function $U$ with a regular $(F,\alpha)$-decay has an
integrable long-distance decay if and only if $\alpha > 2$.
It has a sub-Gaussian decay if and only if 
$\lim_{t \to \infty} F(t)/$\hspace{2pt}$\sqrt{\log t} = \infty$ or, equivalently,
$\lim_{t \to \infty} \e^{-\delta t^2} F^{-1}(t)  = 0$ for all $\delta >0$.
Consequently, $\alpha < \infty$ implies sub-Gaussian decay.
\end{nummer}
\end{remarks}
Theorem~\ref{ThmClassical} below deals with
the class of functions having an integrable sub-Gaussian regular decay.
This class is 
illustrated by the following four examples presented in the order
of gradually slower decay.
\begin{examples} \label{Beispiele1}
\hfill
\begin{nummer}
\item
$\displaystyle\quad
 U(x)= g \exp\left[- \frac{|x|^2}{\lambda^2 \log\left(|x|/\mu\right)}(1+o(1))\right] 
$\hfill
as $\quad |x| \to \infty$,\\[1ex]
\noindent
with some constants $g$, $\lambda$, $\mu>0$. Here $F(t)\sim\lambda \sqrt{(\log t) \log (\sqrt{\log t})}\;$
 ($t\to\infty$), $\alpha=\infty$.
\item
$\displaystyle\quad 
U(x)= g \exp\left[-\left(|x|/\lambda\right)^\beta (1+o(1))\right]
$\hfill
as $\quad |x| \to \infty$,\\[1ex]
\noindent
with some constants $0 < \beta < 2$ and $g$, $\lambda > 0$. 
It defines the class of functions with decays 
called \emph{stretched-Gaussian} in \cite{HLW99}.
Here $F(t)\sim \lambda \left(\log t\right)^{1/\beta}\;$ ($t\to\infty$), $\alpha=\infty$. 
\item
$\displaystyle\quad
U(x)= \frac{g_0}{|x|^{\alpha}}(1+o(1))
$\hfill
as $\quad |x| \to \infty$,\\[1ex]
\noindent
with some constant $g_0> 0$ and $\alpha > 2$. It  defines the class of functions with
\emph{integrable algebraic decay}. 
Here $F(t)\sim \left(g_0 t\right)^{1/\alpha}\;$  ($t\to\infty$).   
\item
$\displaystyle\quad
U(x)= \frac{g_0}{|x|^{\alpha}}\left|\log\left(|x|/\mu\right)\right|(1+o(1))
$\hfill
as $\quad |x| \to \infty$,\\[1ex]
\noindent
with some constants $g_0$, $\mu > 0$ and $\alpha > 2$. Here
$F(t) \sim (g_0 t)^{1/\alpha}\left[\left(\log t\right)/\alpha\right]^{1/\alpha}\;$  ($t\to\infty$).
\end{nummer}
\end{examples}
Remarkably, it is possible to determine rather explicitly the Lifshits tail caused 
by any impurity potential which shares the common decay properties of Examples~\ref{Beispiele1}.
\begin{theorem}\label{ThmClassical}
Let the positive impurity potential $U$ be in ${\rm L}^2_{\mbox{\rm \scriptsize loc}}(\rz^2)$ 
and have a regular $(F,\alpha)$-decay with $\alpha \in ]2, \infty]$. 
Furthermore, let  $U$ have sub-Gaussian decay (\ref{SubGauss}).
Then the integrated density of states falls off to zero at $\varepsilon_0>0$
asymptotically according to
\begin{equation}\label{eq:classical}
\log N(\varepsilon_0 + E) \sim - C(\alpha, \varrho) \, E^{2/(2 - \alpha)}\, 
f^{\#}(E^{\alpha/(2 - \alpha)})  \qquad (E \downarrow 0).
\end{equation} 
Here $C(\alpha, \varrho):= \frac{\alpha -2}{2} 
\left[\frac{2 \pi \varrho}{\alpha} 
\Gamma\left(\frac{\alpha -2}{\alpha}\right)\right]^{\alpha/(\alpha - 2)}$,
$\Gamma$ denotes Euler's gamma function and the function $f^{\#}$ is the de Bruijn 
conjugate of the function 
$f: t \mapsto f(t):= \left[t^{-1/\alpha} F(t)\right]^{2\alpha/(2-\alpha)}$. 
\end{theorem}
\begin{remarks}
\begin{nummer}
\item
For the definition of the de Bruijn conjugate and 
some examples of de Bruijn conjugate pairs,
see Appendix~\ref{AppendixRegVar}.
\item
For the boundary case $\alpha=\infty$ the assertion (\ref{eq:classical}) reduces to
\begin{equation}
\log N(\varepsilon_0 + E) \sim - \pi \varrho \, f^{\#}(E^{-1}) \qquad \quad (E \downarrow 0),
\end{equation}
where $f^{\#}$ is the de Bruijn conjugate of $f: t \mapsto \left[F(t)\right]^{-2}$.
\item
The Lifshits tail (\ref{eq:classical}) neither depends on Planck's constant $2 \pi \hbar$ nor 
on the magnetic field $B$ -- an indication of its classical character. 
In fact, the asymptotic equivalence (\ref{eq:classical}) remains valid if one substitutes
for $N(\varepsilon_0 +E)$  the \emph{classical integrated density of states}
\begin{equation}\label{clIDOS}
N_c(E):=\frac{m}{2\pi \hbar^2}  \int_\Omega \! \d {\mathbb P}(\omega)\, 
\left[E-V_\omega(0)\right]\, \Theta\left(E-V_\omega(0)\right)
\end{equation}
which is obviously independent of $B$; confer \cite{HLW99}. 
The magnetic field (and hence Planck's constant) will show up only in sub-leading corrections to 
(\ref{eq:classical}).
Even so, we stress that in case $\alpha \in [4, \infty]$ the validity of
the leading behaviour (\ref{eq:classical}) already
requires the presence of the magnetic field, see also the next remark.
\item
As was pointed out in \cite{BHKL95}, the Lifshits tail (\ref{eq:classical})
for $\varepsilon_0>0$ coincides with that for vanishing magnetic field, $\varepsilon_0=0$, at the
corresponding spectral bottom in case $U$ has an integrable algebraic decay which 
is slow in that $\alpha\in ]2,4[$,
compare Example~\ref{Beispiele1} (iii) below with \cite{Pas77} or \cite[Corollary 9.14]{PaFi92}.
By following the lines of reasoning for vanishing magnetic field in \cite{Pas77} or \cite{PaFi92}
and comparing the result with (\ref{eq:classical}), it can be seen that this circumstance occurs for
every $U$ with some regular $(F,\alpha)$-decay provided that $\alpha\in ]2,4[$. 
\end{nummer} 
\end{remarks}
We now return to the Examples~\ref{Beispiele1}. According to Theorem~\ref{ThmClassical} the
Lifshits tails caused by these impurity potentials turn out as follows.
\subsection*{Examples~\ref{Beispiele1} (revisited)}
\begin{nummer}
\item
$\quad\displaystyle
\log N(\varepsilon_0 + E) \sim - \pi \varrho \, \lambda^2 \left| \log E \right| 
 \; \log (\left| \log E \right|^{1/2})
$\hfill 
($E\downarrow 0$),
\item
$\quad\displaystyle
\log N(\varepsilon_0 + E) \sim - \pi \varrho \, \lambda^2 \left| \log E \right|^{2/\beta}
$\hfill 
($E\downarrow 0$),
\item 
$\quad\displaystyle \log N(\varepsilon_0 + E) \sim - C(\alpha, \varrho) \, 
\left(\frac{g_0}{E}\right)^{2/(\alpha-2)}
$\hfill  
($\displaystyle E\downarrow 0$),
\item
$\quad \displaystyle
\log N(\varepsilon_0 + E) \sim - C(\alpha, \varrho) \, \left(\frac{g_0}{E}\right)^{2/(\alpha-2)} \,
\left(\frac{|\log E |}{\alpha-2}\right)^{2/(\alpha-2)}
$\hfill 
($E \downarrow 0$).
\end{nummer}
\hfill
\begin{remark}
To our knowledge, the Lifshits tails (ii) and (iii) were first presented and proven
in \cite{HLW99} and \cite{BHKL95},
respectively. In this sense, Theorem~\ref{ThmClassical} unifies and generalizes these previous results.
For the derivation of (i) and (iv) we took advantage of the relation 
$f^{\#}(t) \sim 1/f(t)\;$ ($t\to \infty$), valid in both cases, see Appendix~\ref{AppendixRegVar}.
The examples nicely illustrate the fact:
the slower the long-distance decay of $U$, the faster the low-energy fall-off of $N$.
\end{remark}
Theorem~\ref{ThmClassical} covers many but not all impurity potentials
$U$ with sub-Gaussian (and integrable) decay, confer Remark~\ref{nummer1} (iii).
Unfortunately, we do not know of a general theory which determines the Lifshits tails caused by the
remaining potentials. Our methods fail in general. 
\subsection{Classical versus quantum Lifshits tails}
An interesting question is which long-distance decay of the
impurity potential $U$ causes a quantum Lifshits tail, in the sense
that the leading fall-off of $N(\varepsilon_0+E)$ does
not coincide with that of $N_c(E)$ for $E\downarrow 0$. 
Theorem~\ref{ThmQuantum} and Theorem~\ref{ThmClassical} give a partial answer.
By passing from regular sub-Gaussian to super-Gaussian decay of $U$,
the corresponding Lifshits tail changes from classical to quantum.
Along such a route, Gaussian decay discriminates
between classical and quantum Lifshits tailing \cite{HLW99}.
In case $U$ has sub-Gaussian but no regular decay, we do not know
whether it causes a classical or quantum Lifshits tail.
But we conjecture that the Lifshits tail of $N$ can be universally deduced from the 
lower bound in the subsequently given inequalities (\ref{bounds}). If this is true, genuine
quantum effects in this tail should emerge for certain $U$ having sub-Gaussian but no regular decay.
For example, if $U$ oscillates up to infinity between two functions with
different regular sub-Gaussian decay (confer the second example in Remark~\ref{nummer1} (iii)), 
the oscillation length has to compete with the magnetic length in the lower bound 
in (\ref{bounds}) through the convolution.
Clearly, the Golden-Thompson type of upper bound in (\ref{bounds}) is not sharp
enough to prove the above conjecture.
\section{Proof of Theorem~\ref{ThmClassical}}\label{Proof}
For the proof of Theorem~\ref{ThmClassical} we follow exactly the
strategy in \cite{HLW99} and \cite{BHKL95}, which in turn follow the strategy in \cite{Pas77}.
We note that the assumptions of the theorem imply Assumption~\ref{Genass}. 
The Tauberian theorem in Appendix~\ref{AppendixTauber} 
(with $\eta=\varepsilon_0$ and $\gamma=2/\alpha$) 
shows that the claimed low-energy
fall-off of $N$ is equivalent to the leading asymptotic fall-off
\begin{equation}
\lim_{t \to \infty}\, \left[F(t)\right]^{-2} \,  \log \widetilde N(t) = - \pi \varrho \,
\Gamma\left(\frac{\alpha -2}{\alpha}\right) 
\end{equation}
of its shifted Laplace-Stieltjes transform $\widetilde N(t) := 
\int_0^\infty \! \d N(\varepsilon_0 +E) \, \e^{-tE}$, for long ``time'' $t>0$.
To determine the long-time behaviour of $\widetilde N$ we use the pointwise
sandwiching bounds
\begin{multline}\label{bounds}
\frac{1}{2 \pi \ell^2} \exp\left[- \varrho \intr[2]{x} 
\left( 1 - \e^{-t \left(|\phi_0|^2*U\right)(x)}\right)\right] \\
\leq  \widetilde N (t) 
\leq \frac{\e^{t\varepsilon_0}}{4 \pi \ell^2 \sinh(t\varepsilon_0)} \,
\exp\left[ - \varrho \intr[2]{x}\left(1 - \e^{-t U(x)}\right)\right].
\end{multline}
They rely on a Jensen-Peierls and Golden-Thompson type of inequality and are
proven in \cite{HLW99} and \cite{BHKL95}. 
Here
\begin{equation}
\left(|\phi_0|^2*U\right)(x) := \frac{1}{2 \pi \ell^2} 
\intr[2]{y}  e^{- |x-y|^2/2\ell^2} \, U(y)
\end{equation}
denotes the Lebesgue convolution of $U$ and the Gaussian probability density 
\linebreak 
$|\phi_0(x)|^2:=\exp\left[-|x|^2/(2\ell^2)\right]/(2\pi \ell^2)$.

We proceed  
by deducing the long-time fall-offs of the
lower and upper bound in (\ref{bounds}) from the long-distance decays
of $|\phi_0|^2*U$ and $U$, respectively. This is accomplished by 
the following 
\begin{lemma}\label{Abfall}
Let $W$ be a positive integrable function on $\rz^2$ having a regular 
$(F,\alpha)$-decay with $\alpha \in ]2, \infty]$. Then
\begin{equation}\label{eq:Abfall}
\lim_{t \to \infty} \, \left[F(t)\right]^{-2} \,
\intr[2]{x} \left(1 - \e^{-tW(x)}\right) =
\intr[2]{x} \left(1 - \e^{-|x|^{-\alpha}}\right) 
= \pi\, \Gamma\left(\frac{\alpha -2}{\alpha}\right).
\end{equation} 
Here we employ the conventions (\ref{convention}) to deal 
with the boundary case $\alpha = \infty$ simultaneously.
\end{lemma}
The proof of Theorem~\ref{ThmClassical} is then completed by 
showing that $|\phi_0|^2*U$ has the same 
sub-Gaussian, regular $(F,\alpha)$-decay as $U$. This is the content of
\begin{lemma}\label{Faltung}
Let $U$ be a positive integrable function on $\rz^2$.  
If $U$ has a sub-Gaussian, regular $(F,\alpha)$-decay with $\alpha \in ]2, \infty]$,
then
the convolution $|\phi_0|^2*U$ has the same sub-Gaussian, 
regular $(F,\alpha)$-decay.  
\end{lemma}
Basically, the proofs of both lemmata follow the proofs
of Lemma 3.4 and Lemma 3.5 in \cite{HLW99}. The details are as follows.
\begin{proof}[Proof of Lemma~\ref{Abfall}]
The substitution $x =: F(t) \, 
\xi$ in the left integral in (\ref{eq:Abfall}) yields
\begin{equation}
\intr[2]{x} \left(1 - \e^{-tW(x)}\right) = \left[F(t)\right]^2 
\intr[2]{\xi} \left(1 - \e^{-tW(F(t)\xi)}\right).
\end{equation}
Using the inverse $F^{-1}$ of $F$ and the fact that $W$ has regular 
$(F,\alpha)$-decay (\ref{AbfallU}), one shows that 
for every $\varepsilon \in ]0,1[$ there exists $T_\varepsilon >0$ such that
\begin{equation}\label{Schranken}
\frac{1}{F^{-1}\left((1+\varepsilon)F(t)|\xi|\right)} \leq W(F(t)\xi) \leq
\frac{1}{F^{-1}\left((1-\varepsilon)F(t)|\xi|\right)}
\end{equation}
for all $t > T_\varepsilon$. Since $F^{-1} \in R_\alpha$, that is, 
$\lim_{t \to \infty} 
F^{-1}(t)/F^{-1}(t |\xi|) = |\xi|^{-\alpha}$, $|\xi|>0$, 
the bounds (\ref{Schranken}) imply
\begin{equation}
\lim_{t \to \infty} t \, W(F(t)\xi) = |\xi|^{-\alpha}, \qquad 0 < |\xi| \neq 1.
\end{equation}
The claimed result now follows by interchanging limit and integration by 
applying the dominated-convergence theorem.
In order to show that this theorem is indeed
applicable we have to distinguish the cases $\alpha<\infty$ and $\alpha = \infty$.
In the first case, we may use the decomposition (\ref{eq:repRegVar})
together with Proposition~\ref{Potter} to further estimate (\ref{Schranken}) and 
construct an upper bound on $t W(F(t) \xi)$ which is independent of $t$ and 
has an integrable long-distance decay. 
In the second case, we use (\ref{schnelleswachstum}) instead of (\ref{eq:repRegVar})
and Proposition~\ref{Potter} in (\ref{Schranken}).
\end{proof}
\begin{proof}[Proof of Lemma~\ref{Faltung}]
As in the proof of Lemma 3.5 of \cite{HLW99} we construct asymptotically coinciding
upper and lower bounds on $|\phi_0|^2 * U$. For the upper bound we pick 
$\varepsilon \in ]0,1[$ and split the convolution integral into two integrals with
domains of integration inside and outside a disk with radius
$\varepsilon |x|$
centered about the origin and estimate the two parts separately as follows
\begin{multline}
\int_{|y|\leq\varepsilon |x|}\mkern-10mu \d^2 y \, \,
       U(x-y)\,\left|\phi_0(y)\right|^2  \leq 
        \sup_{|y|\leq \varepsilon |x|} U(x-y)  \\
   \leq \sup_{|y|\leq \varepsilon |x|} 
            \frac{1}{F^{-1}\left((1- \varepsilon) |x-y|\right)} 
    \leq  \frac{1}{F^{-1}\left((1- \varepsilon)^2 |x|\right)}, \quad
\end{multline}
for sufficiently large $|x|$, since $U$ has a regular $(F,\alpha)$-decay and $F^{-1}$ is
strictly increasing. Moreover, estimating the Gaussian $\left|\phi_0\right|^2$
on the domain of integration yields
\begin{equation}
\int_{|y|>\varepsilon |x|}\mkern-10mu \d^2 y \, 
    U(x-y) \, \left|\phi_0(y)\right|^2 \leq  
         \frac{1}{2\pi \ell^2} \, \e^{-\varepsilon^2 |x|^2/2 \ell^2}\,
         \intr[2]{y} U(y).
\end{equation}
Since $U$ has sub-Gaussian decay it follows that
$\lim_{t \to \infty} F^{-1}(t) \,\e^{-\delta t^2} = 0$ for all $\delta > 0$
such that the first term dominates the asymptotics of $|\phi_0|^2 * U$. 
Employing the facts that $F$ is strictly increasing and 
in $R_{1/\alpha}$, we therefore arrive at
\begin{equation}
\liminf_{|x| \to \infty} \, \frac{F\left(1/(|\phi_0|^2 * U)(x)\right)}{|x|} \geq
\frac{(1- \varepsilon)^2}{(1+ \varepsilon)^{1/\alpha}}.
\end{equation}
For a lower bound
we may proceed similarly
\begin{multline}
\left(|\phi_0|^2 * U\right)(x) \geq 
\int_{|y| \leq \varepsilon |x|}\mkern-10mu \d^2 y \, \, U(x-y) \left|\phi_0(y)\right|^2 \\
\geq \inf_{|y|\leq \varepsilon |x|} U(x-y) \, 
\int_{|z| \leq \varepsilon |x|} \mkern-10mu \d^2 z \, \left|\phi_0(z)\right|^2 
\geq \frac{1- \varepsilon}{F^{-1}\left((1+ \varepsilon)^2 |x|\right)}, \quad
\end{multline}
for sufficiently large $|x|$, which gives
\begin{equation}
\limsup_{|x| \to \infty} \, \frac{F\left(1/(|\phi_0|^2 * U)(x)\right)}{|x|} \leq
\frac{(1+\varepsilon)^2}{(1- \varepsilon)^{1/\alpha}}.
\end{equation}
This completes the proof since $\varepsilon$ may be picked arbitrarily small.
\end{proof}
\appendix
\section{Elementary facts about ``Regular Variation''}\label{AppendixRegVar}
The theory of regular variation was initiated by Jovan Karamata in 1930.
For the proofs of the properties quoted below
and many further related results we refer to the excellent monograph \cite{BGT89}.
We recall that
  two functions $F$ and $G$ are \emph{asymptotically equivalent},
  in symbols $F(t)\sim G(t)\quad (t\to\infty)$, if
   $F(t)/G(t)\to 1$ as $t\to\infty$.
  
\subsection*{Slow variation.}
A positive, measurable function $f$ on the positive half-line is said to be
  \emph{slowly varying} (at infinity) if
  $ f(ct)/f(t) \to 1 $ as $t\to\infty$ holds for all $c>0$,
in symbols $f\in R_0$.
Standard examples of slowly varying functions are
\begin{equation}\label{eq:ExaSloVar}
 t\mapsto a_0\prod_{j=1}^{n}\left[\log_j(t)\right]^{a_j}
 \quad\mbox{and}\quad
 t\mapsto \exp\left[(\log t)^a\right]
\end{equation}
 where $a_j\in \rz$, $a\in]0,1[$,
 and $\log_j$ denotes the $j$-times iterated logarithm.
In fact, for every $f\in R_0$ there is an arbitrarily often
differentiable function $f_0$ with $f_0(t) \sim f(t)$.
If in addition $f$ is monotone, $f_0$ can be chosen monotone, too.
Even so, not every $f\in R_0$ is equivalent to a monotone function.
Actually, there are $f\in R_0$ with $\liminf_{t\to\infty} f(t)=0$ and
  $\limsup_{t\to\infty} f(t)=\infty$.
Nevertheless, the rate of growth (or decay) is bounded by any (inverse) power
  according to so-called Potter bounds \cite[Theorem 1.5.6]{BGT89}
  \begin{proposition}\label{Potter}
    Let $f\in R_0$, then for any pair of constants
    $A>1$ and  $\delta>0$ there exists $T\in\rz$,
    possibly depending on $A$ and $\delta$, such that
    \begin{equation}
      f(t) / f(s) \leq
      A \max\left\{(t/s)^\delta, (t/s)^{-\delta}\right\}
    \end{equation}
    for all $t, s \geq T$.
  \end{proposition}

\subsection*{The de Bruijn conjugate.}
For $f\in R_0$ there exists
  $f^{\#}\in R_0$, unique up to asymptotic equivalence, such that
  \begin{equation}\label{eq:DefDeBrujnConj}
    f(t)\, f^{\#}(t f(t) ) \to 1,\quad
    f^{\#}(t)\, f(t f^{\#}(t) ) \to 1\quad
    \mbox{as}\quad t\to\infty,
  \end{equation}
  and $f^{\#\#}\sim f$.  
The function $f^{\#}$ is called the \emph{de Bruijn conjugate} of $f$ and
  $(f, f^{\#})$ is referred to as a \emph{conjugate pair}.
With positive constants $A, B, \beta>0$ each of the following three pairs is a
  conjugate pair:
  \begin{equation}
 (f(At)  , f^{\#}(B t)), \qquad
      (A f(t) , A^{-1}f^{\#}(t)),\qquad
      ([f(t^\beta)]^{1/\beta} ,[f^{\#}(t^\beta)]^{1/\beta}).
    \end{equation}
In some cases, the de Bruijn conjugate can be calculated explicitly,
  confer \cite[Appendix~5]{BGT89}.
Notably, if $f(t f(t))\sim f(t)\quad(t \to \infty)$ then
  $f^{\#}(t)\sim 1/f(t)$.
This simple criterion applies, for instance, to the first example in~\eqref{eq:ExaSloVar}
  and gives in particular that
  $([\log t]^{\beta},[\log t]^{-\beta})$ is a conjugate pair
  for any real $\beta\neq 0$.
For the de Bruijn conjugate of the second example in \eqref{eq:ExaSloVar}
  there are also explicit expressions available
  which are somewhat complicated if $a\geq1/2$.

\subsection*{Regular variation} 
A positive measurable function $F$ is said to be \emph{regularly varying}
  (at infinity) if 
  $ \lim_{t\to\infty}F(ct)/F(t)\in ]0,\infty[$
  for all $c>0$ in a set of strictly positive Lebesgue measure.
Then there is $\gamma \in \rz$ such that
\begin{equation}\label{defregvar}
  \lim_{t\to\infty}\,\frac{F(c t)}{F(t)} = c^\gamma, 
\end{equation}  
  for all $c>0$.
We call such a $F$ \emph{regularly varying of index $\gamma$} and write $F\in R_\gamma$.
Every $F\in R_\gamma$ has the form
\begin{equation}\label{eq:repRegVar}
  F(t)=t^\gamma f(t)  
\end{equation}
  with some $f \in R_0$.
For $\gamma\in\rz$, $\delta > 0$, 
$F\in R_\gamma$, and $G\in R_{\delta}$ one has 
  $F(G(\cdot)) \in R_{\gamma \delta}$.
Every $F \in R_\gamma$ with $\gamma \neq 0$
  is asymptotically equivalent to a monotone function.
Its inverse belongs to $R_{1/\gamma}$.
More explicitly, if $F(t)\sim t^{\gamma \delta} (f(t^\delta))^\gamma$ with
  some $f \in R_0$ and $\gamma, \delta > 0$
  and $G$ is an \emph{asymptotic inverse} of $F$, that is, 
  $G(F(t)) \sim F(G(t)) \sim t$, then
  $G(t)\sim t^{1/(\gamma \delta)} (f^{\#}(t^{1/\gamma}))^{1/\delta}$.

\subsection*{Rapid variation}
The boundary cases $\gamma = \pm \infty$ in (\ref{defregvar}) lead to the
notion of rapidly varying functions, where we adopt the conventions
\begin{multline}\label{convention}
\quad c^\infty := \left\{\begin{array}{lcc}
        0 & & 0 < c < 1 \\
        1 & \mbox{if} & c=1   \\
        \infty & & c > 1 
        \end{array}\right. 
\quad\mbox{and}\quad
c^{-\infty} := \left\{\begin{array}{lcc}
      \infty & & 0 < c < 1 \\
       1 & \mbox{if} & c=1   \\
       0 & & c > 1.
      \end{array}\right.   
\end{multline}
More precisely, a positive measurable function $F$ is said to be 
\emph{rapidly varying of index $\pm \infty$} if (\ref{defregvar})
holds with $\gamma = \pm \infty$ for all $c>0$, in symbols $F\in R_{\pm\infty}$.
If $F\in R_{\infty}$ is non-decreasing, then for 
any $A < 1$ and $\gamma \in \rz$ there exists $T>0$ such that
\begin{equation}\label{schnelleswachstum}
\frac{F(ct)}{F(t)} \geq A c^\gamma
\end{equation}
for all $t>T$ and $c\geq 1$. 
Moreover, if $F \in R_0$ with $F(t) \to \infty$ as $t \to \infty$
and $G$ is an asymptotic inverse of $F$, then $G \in R_\infty$.
\section{A Tauberian theorem of exponential type}\label{AppendixTauber}
%
\begin{theorem}\label{Taubertheorem}
  Let $N$ be the distribution function of a positive Borel measure
  on the real line $\rz$.
  Assume there is a constant $\eta \in \rz$ such that $N(E)=0$
  for all $E \leq \eta$. Moreover, define the shifted Laplace-Stieltjes transform
  of $N$ by
  \begin{equation}\label{eq:TLaplace}
     \widetilde N (t) := \int_0^\infty \! \d N(\eta + E) \, \e^{-t E} = 
     \e^{t \eta} \int_\eta^\infty \! \d N(E) \, \e^{-t E} ,
    \quad t > 0 ,
  \end{equation}
  and suppose that $\widetilde N (\tau) < \infty$ for some $\tau > 0$. 
  Let $(f,f^{\#})$ be a conjugated pair of slowly varying functions 
  and $\gamma \in [0,1[$. Then
  \begin{equation}\label{eq:TTLT}
    \log \widetilde N (t) \, \sim  \, - t^\gamma \left[f(t)\right]^{\gamma - 1}
    \qquad \quad (t \to \infty),
    \end{equation}
if and only if
  \begin{equation}\label{eq:TTIDOS}
    \log  N (\eta +E) \, \sim \, - (1 - \gamma) 
    \left(\frac{\gamma}{E}\right)^{\gamma/(1-\gamma)} 
      f^{\#}(E^{1/(\gamma -1)}) \qquad (E \downarrow 0).
  \end{equation}
\end{theorem}
\begin{remarks}
\begin{nummer}
\item
For $\gamma > 0$ the theorem is due to de Bruijn as one may check by 
setting $A=1$, $B=(\gamma-1)/\gamma$, $\beta=\gamma/(\gamma-1)$, and
$L(E)=\gamma^{\gamma/(1-\gamma)}$\hspace{2pt}$\left[f^{\#}(E^{1/\gamma})\right]^\gamma$
in \cite[Theorem 2]{B59}, see also \cite[Theorem 4.12.9]{BGT89}. It was 
re-discovered in a slightly different formulation by Minlos and 
Povzner \cite[Appendix]{MiPo67}, see also \cite[Theorem 9.7]{PaFi92}. 
\item
For the boundary case $\gamma=0$ the assertion of the theorem reduces to
\begin{equation}\label{Tauberlangsam}
\lim_{t \to \infty} f(t) \log \widetilde N(t) = -1 \qquad\mbox{if and only if}\qquad
\lim_{E \downarrow 0} \frac{\log N(\eta +E)}{f^{\#}(1/E)} = -1.
\end{equation}
The equivalence (\ref{Tauberlangsam})
was proven in \cite{HLW99} only for 
$f(t)=C \left(\log t\right)^{-\beta}$, $C>0$, $\beta \geq 1$. For 
$f(t)=C \left(\log t\right)^{-1}$ it is a corollary of one of Karamata's early
results \cite{Kar31}.
\item
Integrating by parts in (\ref{eq:TLaplace}) gives
\begin{equation}\label{Npartiell}
\widetilde N (t) = t \int_\eta^\infty \! \d E \,  N(E) \,\e^{-t (E-\eta)},
\quad t > \tau .
\end{equation}
\end{nummer}
\end{remarks}
\begin{proof}[Proof of Theorem~\ref{Taubertheorem} for $\gamma=0$]
We will only give an outline since the proof 
copies exactly the strategy of \cite{P61}. First note that the theorem is 
immediate if $f(t) \to c > 0$ as $t \to \infty$ since 
$\lim_{E\downarrow 0} N(\eta + E) = \lim_{t \to \infty} \widetilde N (t)$. 
We will therefore assume  
throughout the rest $f(t) \to 0$ ($f^{\#}(t) \to \infty$)
as $t \to \infty$  and, moreover, $\eta =0$ without loss of generality. 
Using Lemma~\ref{TL1} below one shows that for every
$\varepsilon > 0$ 
\begin{equation}
\limsup_{E \downarrow 0} 
\frac{\log N(E)}{f^{\#}(1/E)} \leq -1 + \varepsilon \quad \mbox{and} \quad
\liminf_{t \to \infty} f(t) \log \widetilde N(t) \geq -1 - \varepsilon 
\end{equation}
provided that (\ref{eq:TTLT}) and (\ref{eq:TTIDOS}) holds, respectively.
To complete the proof we note that Lemma~\ref{TL2} below gives 
\begin{equation}
\liminf_{E \downarrow 0} 
\frac{\log N(E)}{f^{\#}(1/E)} \geq -1  \quad \mbox{and} \quad
\limsup_{t \to \infty} f(t) \log \widetilde N(t) \leq -1,
\end{equation}  
again supposing that (\ref{eq:TTLT}) and (\ref{eq:TTIDOS}) 
holds, respectively.
\end{proof}
\begin{lemma}\label{TL1}
In the setting of Theorem~\ref{Taubertheorem} assume $\gamma=\eta=0$. 
Then for every $\varepsilon >0$ and $E > 0$
\begin{equation}
N(E) \leq \e^{\varepsilon f^{\#}(1/E)} \, \widetilde 
N\left(\varepsilon\frac{f^{\#}(1/E)}{E}\right).
\end{equation}
\end{lemma}
\begin{proof}
Since $N(E) \leq \e^{tE} \widetilde N(t)$ for all $t >0$ (compare 
\cite[Equation (A.4)]{HLW99}) the inequality follows by choosing 
$t=\varepsilon \, f^{\#}(1/E) / E$.
\end{proof}
\begin{lemma}\label{TL2}
In the setting of Theorem~\ref{Taubertheorem} assume $\gamma=\eta=0$ and 
that either (\ref{eq:TTLT})
or (\ref{eq:TTIDOS}) holds.
If furthermore $f(t) \to 0$ as $t \to \infty$, then
\begin{equation}
N(E) \geq \widetilde N\left(\frac{f^{\#}(1/E)}{E}\right) - 2 \exp\left(-\frac{9}{8}f^{\#}(1/E)\right) 
\end{equation} 
for sufficiently small $E>0$.
\end{lemma}
\begin{proof}
We put 
$t_E := f^{\#}(1/E)/E$ and split the domain of integration in 
(\ref{Npartiell}) into the three parts $[0,E]$, $[E,2E]$,
and $[2E,\infty[$. The integral over the first part is estimated 
according to
\begin{equation}
t_E \int_0^E \! \d u \,  N(u) \e^{-t_E u} \leq N(E) 
\end{equation}
due to monotonicity of $N$.
We now employ Lemma~\ref{TL1} or (\ref{eq:TTIDOS}) directly to 
show that $N(E) \leq \exp\left(- f^{\#}(1/E)/2\right)$ 
for sufficiently small $E>0$
which implies the following upper bound for the integral over the 
second part
\begin{multline}
t_E \int_E^{2E} \! \d u \,  N(u) \, \e^{-t_E u} \leq
t_E \int_E^{2E} \! \d u \, \exp\left[ - t_E u
  \left(1 + 
    \frac{E}{2u}\frac{f^{\#}(1/u)}{f^{\#}(1/E)}\right)\right] \\
 \leq  t_E \int_E^{2E} \! \d u \, \exp\left(-\frac{9}{8} t_E u\right) \leq 
\exp\left(-\frac{9}{8} f^{\#}(1/E)\right).
\end{multline}
Here we used Proposition~\ref{Potter} with $A=\sqrt{2}$ and  $\delta=1$ which yields
$f^{\#}(1/u)/ u \geq f^{\#}(1/2E)/(2\sqrt{2}E)$\hspace{2pt}$\geq  f^{\#}(1/E))/(4E)$ 
for sufficiently small $E>0$, since $u \in [E,2E]$ and $f^{\#}\in R_0$. 
Finally, 
\begin{multline}
t_E \int_{2E}^\infty \! \d u \,  N(u) \e^{-t_E u} \leq
\exp\left(-\frac{3}{2} f^{\#}(1/E)\right) \, t_E \int_{2E}^\infty \! \d u \,  N(u) \, \e^{-t_E u/4} 
\\
\leq  4 \widetilde N\left(\frac{t_E}{4}\right) \, \exp\left(-\frac{3}{2} f^{\#}(1/E)\right)  
\leq \exp\left(-\frac{9}{8} f^{\#}(1/E)\right)  
\end{multline}
for sufficiently small $E>0$, for which 
$\widetilde N(t_E /4) < \widetilde N(\tau) < \infty$ and $f^{\#}(1/E) \to \infty$ as $E\downarrow 0$.
\end{proof}

%
%
\section*{Acknowledgements}
Our thanks go to L\'aszl\'o Erd\H{o}s for stimulating part of this work.
We are also much indebted to Alexander Bendikov and Charles M. Goldie for helpful
remarks in relation to the Tauberian theorem.
This work was supported by the Deutsche Forschungsgemeinschaft.
\bibliographystyle{amsalpha}

\end{document}

%% file: test2.pictex.tex
\font\thinlinefont=cmr5
\begingroup\makeatletter\ifx\SetFigFont\undefined%
\gdef\SetFigFont#1#2#3#4#5{%
  \reset@font\fontsize{#1}{#2pt}%
  \fontfamily{#3}\fontseries{#4}\fontshape{#5}%
  \selectfont}%
\fi\endgroup%
\mbox{
\hspace{0.7cm}
\beginpicture
\setcoordinatesystem units <0.4cm,0.28cm> 
\unitlength=0.52493cm
\linethickness=1pt
\setplotsymbol ({\makebox(0,0)[l]{\tencirc\symbol{'160}}})
\setshadesymbol ({\thinlinefont .})
\setlinear
%
\linethickness= 0.500pt
\setplotsymbol ({\thinlinefont .})
\special{ps: gsave 0 0 0 setrgbcolor}\putrule from  2.667 11.430 to 25.908 11.430
%
%
\plot 25.654 11.366 25.908 11.430 25.654 11.494 /
\special{ps: grestore}%
%
\linethickness= 0.500pt
\setplotsymbol ({\thinlinefont .})
\special{ps: gsave 0 0 0 setrgbcolor}\putrule from  4.763 10.096 to  4.763 25.432
%
%
\plot  4.826 25.178  4.763 25.432  4.699 25.178 /
\special{ps: grestore}%
%
\linethickness= 0.500pt
\setplotsymbol ({\thinlinefont .})
\special{ps: gsave 0 0 0 setrgbcolor}\putrule from  4.572 16.192 to  4.953 16.192
\special{ps: grestore}%
%
\linethickness= 0.500pt
\setplotsymbol ({\thinlinefont .})
\special{ps: gsave 0 0 0 setrgbcolor}\putrule from  4.572 21.050 to  4.953 21.050
\special{ps: grestore}%
%
\linethickness= 0.500pt
\setplotsymbol ({\thinlinefont .})
\special{ps: gsave 0 0 0 setrgbcolor}\putrule from  2.667 11.430 to 25.908 11.430
%
%
\plot 25.654 11.366 25.908 11.430 25.654 11.494 /
\special{ps: grestore}%
%
\linethickness= 0.500pt
\setplotsymbol ({\thinlinefont .})
\special{ps: gsave 0 0 0 setrgbcolor}\putrule from  8.572 11.620 to  8.572 11.239
\special{ps: grestore}%
%
\linethickness= 0.500pt
\setplotsymbol ({\thinlinefont .})
\special{ps: gsave 0 0 0 setrgbcolor}\putrule from 16.192 11.620 to 16.192 11.239
\special{ps: grestore}%
%
\linethickness= 0.500pt
\setplotsymbol ({\thinlinefont .})
\special{ps: gsave 0 0 0 setrgbcolor}\putrule from 23.812 11.620 to 23.812 11.239
\special{ps: grestore}%
%
\linethickness= 0.500pt
\setplotsymbol ({\thinlinefont .})
\setdashes < 0.2cm>     
\special{ps: gsave 0 0 0 setrgbcolor}\plot  8.572 11.430  8.572 16.192 /
\plot  8.572 16.192 16.192 16.192 /
\plot 16.192 16.192 16.192 20.955 /
\plot 16.192 20.955 23.812 20.955 /
\plot 23.812 20.955 23.812 23.527 /
\special{ps: grestore}%
%
\linethickness= 0.500pt
\setplotsymbol ({\thinlinefont .})
\setsolid
\special{ps: gsave 0 0 0 setrgbcolor}\putrule from  8.636 11.398 to  8.636 11.402
\plot  8.636 11.402  8.638 11.415 /
\plot  8.638 11.415  8.640 11.434 /
\plot  8.640 11.434  8.642 11.466 /
\plot  8.642 11.466  8.647 11.510 /
\plot  8.647 11.510  8.653 11.565 /
\plot  8.653 11.565  8.659 11.635 /
\plot  8.659 11.635  8.668 11.714 /
\plot  8.668 11.714  8.678 11.803 /
\plot  8.678 11.803  8.689 11.898 /
\plot  8.689 11.898  8.700 11.999 /
\plot  8.700 11.999  8.712 12.103 /
\plot  8.712 12.103  8.723 12.207 /
\plot  8.723 12.207  8.735 12.308 /
\plot  8.735 12.308  8.748 12.408 /
\plot  8.748 12.408  8.761 12.503 /
\plot  8.761 12.503  8.771 12.596 /
\plot  8.771 12.596  8.784 12.683 /
\plot  8.784 12.683  8.797 12.766 /
\plot  8.797 12.766  8.807 12.842 /
\plot  8.807 12.842  8.820 12.916 /
\plot  8.820 12.916  8.831 12.984 /
\plot  8.831 12.984  8.843 13.049 /
\plot  8.843 13.049  8.856 13.109 /
\plot  8.856 13.109  8.867 13.168 /
\plot  8.867 13.168  8.882 13.223 /
\plot  8.882 13.223  8.894 13.276 /
\plot  8.894 13.276  8.907 13.327 /
\plot  8.907 13.327  8.922 13.377 /
\plot  8.922 13.377  8.937 13.424 /
\plot  8.937 13.424  8.951 13.468 /
\plot  8.951 13.468  8.966 13.513 /
\plot  8.966 13.513  8.983 13.557 /
\plot  8.983 13.557  9.002 13.602 /
\plot  9.002 13.602  9.021 13.646 /
\plot  9.021 13.646  9.042 13.688 /
\plot  9.042 13.688  9.068 13.733 /
\plot  9.068 13.733  9.093 13.775 /
\plot  9.093 13.775  9.123 13.818 /
\plot  9.123 13.818  9.155 13.862 /
\plot  9.155 13.862  9.188 13.904 /
\plot  9.188 13.904  9.224 13.947 /
\plot  9.224 13.947  9.267 13.989 /
\plot  9.267 13.989  9.309 14.029 /
\plot  9.309 14.029  9.356 14.072 /
\plot  9.356 14.072  9.406 14.112 /
\plot  9.406 14.112  9.459 14.150 /
\plot  9.459 14.150  9.517 14.190 /
\plot  9.517 14.190  9.578 14.228 /
\plot  9.578 14.228  9.641 14.266 /
\plot  9.641 14.266  9.707 14.302 /
\plot  9.707 14.302  9.779 14.338 /
\plot  9.779 14.338  9.851 14.372 /
\plot  9.851 14.372  9.929 14.406 /
\plot  9.929 14.406 10.010 14.440 /
\plot 10.010 14.440 10.092 14.472 /
\plot 10.092 14.472 10.179 14.503 /
\plot 10.179 14.503 10.270 14.535 /
\plot 10.270 14.535 10.365 14.565 /
\plot 10.365 14.565 10.465 14.597 /
\plot 10.465 14.597 10.569 14.626 /
\plot 10.569 14.626 10.634 14.645 /
\plot 10.634 14.645 10.702 14.662 /
\plot 10.702 14.662 10.772 14.681 /
\plot 10.772 14.681 10.846 14.700 /
\plot 10.846 14.700 10.922 14.719 /
\plot 10.922 14.719 11.000 14.738 /
\plot 11.000 14.738 11.083 14.757 /
\plot 11.083 14.757 11.168 14.776 /
\plot 11.168 14.776 11.256 14.796 /
\plot 11.256 14.796 11.350 14.815 /
\plot 11.350 14.815 11.447 14.836 /
\plot 11.447 14.836 11.549 14.857 /
\plot 11.549 14.857 11.656 14.878 /
\plot 11.656 14.878 11.767 14.899 /
\plot 11.767 14.899 11.883 14.922 /
\plot 11.883 14.922 12.004 14.944 /
\plot 12.004 14.944 12.131 14.967 /
\plot 12.131 14.967 12.262 14.992 /
\plot 12.262 14.992 12.399 15.018 /
\plot 12.399 15.018 12.541 15.043 /
\plot 12.541 15.043 12.689 15.069 /
\plot 12.689 15.069 12.842 15.096 /
\plot 12.842 15.096 13.001 15.124 /
\plot 13.001 15.124 13.164 15.151 /
\plot 13.164 15.151 13.331 15.179 /
\plot 13.331 15.179 13.500 15.208 /
\plot 13.500 15.208 13.676 15.238 /
\plot 13.676 15.238 13.851 15.268 /
\plot 13.851 15.268 14.029 15.297 /
\plot 14.029 15.297 14.207 15.327 /
\plot 14.207 15.327 14.383 15.354 /
\plot 14.383 15.354 14.558 15.384 /
\plot 14.558 15.384 14.730 15.411 /
\plot 14.730 15.411 14.897 15.439 /
\plot 14.897 15.439 15.058 15.466 /
\plot 15.058 15.466 15.212 15.490 /
\plot 15.212 15.490 15.356 15.513 /
\plot 15.356 15.513 15.494 15.536 /
\plot 15.494 15.536 15.617 15.555 /
\plot 15.617 15.555 15.731 15.574 /
\plot 15.731 15.574 15.833 15.589 /
\plot 15.833 15.589 15.922 15.604 /
\plot 15.922 15.604 15.996 15.617 /
\plot 15.996 15.617 16.059 15.627 /
\plot 16.059 15.627 16.112 15.636 /
\plot 16.112 15.636 16.150 15.642 /
\plot 16.150 15.642 16.182 15.646 /
\plot 16.182 15.646 16.201 15.649 /
\plot 16.201 15.649 16.214 15.651 /
\plot 16.214 15.651 16.222 15.653 /
\putrule from 16.222 15.653 to 16.224 15.653
\special{ps: grestore}%
%
\linethickness= 0.500pt
\setplotsymbol ({\thinlinefont .})
\special{ps: gsave 0 0 0 setrgbcolor}\plot 16.288 15.621 16.290 15.625 /
\plot 16.290 15.625 16.292 15.634 /
\plot 16.292 15.634 16.298 15.651 /
\plot 16.298 15.651 16.309 15.676 /
\plot 16.309 15.676 16.322 15.712 /
\plot 16.322 15.712 16.341 15.761 /
\plot 16.341 15.761 16.362 15.818 /
\plot 16.362 15.818 16.389 15.888 /
\plot 16.389 15.888 16.421 15.968 /
\plot 16.421 15.968 16.455 16.057 /
\plot 16.455 16.057 16.493 16.152 /
\plot 16.493 16.152 16.533 16.252 /
\plot 16.533 16.252 16.576 16.355 /
\plot 16.576 16.355 16.618 16.461 /
\plot 16.618 16.461 16.662 16.567 /
\plot 16.662 16.567 16.707 16.671 /
\plot 16.707 16.671 16.749 16.772 /
\plot 16.749 16.772 16.794 16.872 /
\plot 16.794 16.872 16.836 16.967 /
\plot 16.836 16.967 16.878 17.058 /
\plot 16.878 17.058 16.921 17.147 /
\plot 16.921 17.147 16.961 17.230 /
\plot 16.961 17.230 17.001 17.308 /
\plot 17.001 17.308 17.041 17.382 /
\plot 17.041 17.382 17.079 17.454 /
\plot 17.079 17.454 17.117 17.522 /
\plot 17.117 17.522 17.158 17.585 /
\plot 17.158 17.585 17.198 17.647 /
\plot 17.198 17.647 17.236 17.706 /
\plot 17.236 17.706 17.276 17.765 /
\plot 17.276 17.765 17.319 17.820 /
\plot 17.319 17.820 17.361 17.875 /
\plot 17.361 17.875 17.405 17.928 /
\plot 17.405 17.928 17.446 17.979 /
\plot 17.446 17.979 17.490 18.028 /
\plot 17.490 18.028 17.534 18.076 /
\plot 17.534 18.076 17.581 18.125 /
\plot 17.581 18.125 17.630 18.174 /
\plot 17.630 18.174 17.678 18.222 /
\plot 17.678 18.222 17.731 18.269 /
\plot 17.731 18.269 17.784 18.318 /
\plot 17.784 18.318 17.841 18.366 /
\plot 17.841 18.366 17.899 18.413 /
\plot 17.899 18.413 17.960 18.462 /
\plot 17.960 18.462 18.021 18.508 /
\plot 18.021 18.508 18.085 18.555 /
\plot 18.085 18.555 18.150 18.603 /
\plot 18.150 18.603 18.220 18.650 /
\plot 18.220 18.650 18.288 18.694 /
\plot 18.288 18.694 18.360 18.741 /
\plot 18.360 18.741 18.434 18.785 /
\plot 18.434 18.785 18.508 18.830 /
\plot 18.508 18.830 18.584 18.872 /
\plot 18.584 18.872 18.661 18.915 /
\plot 18.661 18.915 18.739 18.957 /
\plot 18.739 18.957 18.817 18.997 /
\plot 18.817 18.997 18.898 19.035 /
\plot 18.898 19.035 18.978 19.073 /
\plot 18.978 19.073 19.058 19.109 /
\plot 19.058 19.109 19.141 19.145 /
\plot 19.141 19.145 19.221 19.181 /
\plot 19.221 19.181 19.304 19.215 /
\plot 19.304 19.215 19.387 19.247 /
\plot 19.387 19.247 19.469 19.279 /
\plot 19.469 19.279 19.552 19.308 /
\plot 19.552 19.308 19.636 19.338 /
\plot 19.636 19.338 19.719 19.365 /
\plot 19.719 19.365 19.806 19.393 /
\plot 19.806 19.393 19.892 19.420 /
\plot 19.892 19.420 19.962 19.442 /
\plot 19.962 19.442 20.036 19.463 /
\plot 20.036 19.463 20.110 19.484 /
\plot 20.110 19.484 20.185 19.505 /
\plot 20.185 19.505 20.263 19.526 /
\plot 20.263 19.526 20.343 19.545 /
\plot 20.343 19.545 20.424 19.566 /
\plot 20.424 19.566 20.508 19.586 /
\plot 20.508 19.586 20.597 19.607 /
\plot 20.597 19.607 20.688 19.626 /
\plot 20.688 19.626 20.784 19.647 /
\plot 20.784 19.647 20.881 19.668 /
\plot 20.881 19.668 20.985 19.689 /
\plot 20.985 19.689 21.090 19.710 /
\plot 21.090 19.710 21.203 19.732 /
\plot 21.203 19.732 21.319 19.753 /
\plot 21.319 19.753 21.440 19.776 /
\plot 21.440 19.776 21.565 19.797 /
\plot 21.565 19.797 21.694 19.820 /
\plot 21.694 19.820 21.827 19.844 /
\plot 21.827 19.844 21.965 19.869 /
\plot 21.965 19.869 22.104 19.892 /
\plot 22.104 19.892 22.248 19.916 /
\plot 22.248 19.916 22.392 19.941 /
\plot 22.392 19.941 22.538 19.964 /
\plot 22.538 19.964 22.682 19.990 /
\plot 22.682 19.990 22.826 20.011 /
\plot 22.826 20.011 22.964 20.034 /
\plot 22.964 20.034 23.099 20.055 /
\plot 23.099 20.055 23.226 20.077 /
\plot 23.226 20.077 23.347 20.096 /
\plot 23.347 20.096 23.459 20.113 /
\plot 23.459 20.113 23.561 20.130 /
\plot 23.561 20.130 23.650 20.144 /
\plot 23.650 20.144 23.728 20.155 /
\plot 23.728 20.155 23.796 20.165 /
\plot 23.796 20.165 23.848 20.174 /
\plot 23.848 20.174 23.893 20.180 /
\plot 23.893 20.180 23.925 20.187 /
\plot 23.925 20.187 23.946 20.189 /
\plot 23.946 20.189 23.961 20.191 /
\plot 23.961 20.191 23.967 20.193 /
\putrule from 23.967 20.193 to 23.971 20.193
\special{ps: grestore}%
%
\linethickness= 0.500pt
\setplotsymbol ({\thinlinefont .})
\special{ps: gsave 0 0 0 setrgbcolor}\plot 24.035 20.161 24.037 20.165 /
\plot 24.037 20.165 24.041 20.176 /
\plot 24.041 20.176 24.052 20.197 /
\plot 24.052 20.197 24.067 20.227 /
\plot 24.067 20.227 24.086 20.269 /
\plot 24.086 20.269 24.111 20.324 /
\plot 24.111 20.324 24.143 20.390 /
\plot 24.143 20.390 24.177 20.464 /
\plot 24.177 20.464 24.215 20.544 /
\plot 24.215 20.544 24.257 20.631 /
\plot 24.257 20.631 24.299 20.720 /
\plot 24.299 20.720 24.344 20.811 /
\plot 24.344 20.811 24.386 20.900 /
\plot 24.386 20.900 24.431 20.987 /
\plot 24.431 20.987 24.471 21.071 /
\plot 24.471 21.071 24.511 21.150 /
\plot 24.511 21.150 24.549 21.226 /
\plot 24.549 21.226 24.585 21.296 /
\plot 24.585 21.296 24.621 21.364 /
\plot 24.621 21.364 24.655 21.425 /
\plot 24.655 21.425 24.687 21.484 /
\plot 24.687 21.484 24.716 21.539 /
\plot 24.716 21.539 24.748 21.592 /
\plot 24.748 21.592 24.778 21.641 /
\plot 24.778 21.641 24.807 21.689 /
\plot 24.807 21.689 24.837 21.736 /
\plot 24.837 21.736 24.864 21.780 /
\putrule from 24.864 21.780 to 24.867 21.780
\plot 24.867 21.780 24.898 21.829 /
\plot 24.898 21.829 24.930 21.876 /
\plot 24.930 21.876 24.964 21.922 /
\plot 24.964 21.922 24.998 21.967 /
\plot 24.998 21.967 25.034 22.013 /
\plot 25.034 22.013 25.072 22.062 /
\plot 25.072 22.062 25.112 22.111 /
\plot 25.112 22.111 25.154 22.159 /
\plot 25.154 22.159 25.199 22.212 /
\plot 25.199 22.212 25.248 22.267 /
\plot 25.248 22.267 25.298 22.322 /
\plot 25.298 22.322 25.351 22.382 /
\plot 25.351 22.382 25.408 22.441 /
\plot 25.408 22.441 25.466 22.502 /
\plot 25.466 22.502 25.523 22.562 /
\plot 25.523 22.562 25.580 22.621 /
\plot 25.580 22.621 25.635 22.678 /
\plot 25.635 22.678 25.684 22.729 /
\plot 25.684 22.729 25.728 22.775 /
\plot 25.728 22.775 25.766 22.813 /
\plot 25.766 22.813 25.796 22.843 /
\plot 25.796 22.843 25.819 22.866 /
\plot 25.819 22.866 25.834 22.881 /
\plot 25.834 22.881 25.840 22.888 /
\plot 25.840 22.888 25.845 22.892 /
\special{ps: grestore}%
%
\linethickness= 0.500pt
\setplotsymbol ({\thinlinefont .})
\special{ps: gsave 0 0 0 setrgbcolor}\putrule from  8.572 11.430 to  8.577 11.430
\putrule from  8.577 11.430 to  8.585 11.430
\plot  8.585 11.430  8.600 11.432 /
\plot  8.600 11.432  8.625 11.434 /
\plot  8.625 11.434  8.661 11.436 /
\plot  8.661 11.436  8.708 11.441 /
\plot  8.708 11.441  8.767 11.445 /
\plot  8.767 11.445  8.837 11.449 /
\plot  8.837 11.449  8.922 11.455 /
\plot  8.922 11.455  9.017 11.464 /
\plot  9.017 11.464  9.123 11.472 /
\plot  9.123 11.472  9.237 11.481 /
\plot  9.237 11.481  9.358 11.489 /
\plot  9.358 11.489  9.485 11.500 /
\plot  9.485 11.500  9.618 11.513 /
\plot  9.618 11.513  9.751 11.523 /
\plot  9.751 11.523  9.887 11.536 /
\plot  9.887 11.536 10.022 11.546 /
\plot 10.022 11.546 10.156 11.559 /
\plot 10.156 11.559 10.287 11.572 /
\plot 10.287 11.572 10.414 11.585 /
\plot 10.414 11.585 10.539 11.597 /
\plot 10.539 11.597 10.657 11.610 /
\plot 10.657 11.610 10.774 11.623 /
\plot 10.774 11.623 10.884 11.635 /
\plot 10.884 11.635 10.988 11.648 /
\plot 10.988 11.648 11.089 11.663 /
\plot 11.089 11.663 11.187 11.676 /
\plot 11.187 11.676 11.278 11.688 /
\plot 11.278 11.688 11.367 11.701 /
\plot 11.367 11.701 11.451 11.716 /
\plot 11.451 11.716 11.532 11.731 /
\plot 11.532 11.731 11.610 11.745 /
\plot 11.610 11.745 11.686 11.760 /
\plot 11.686 11.760 11.758 11.775 /
\plot 11.758 11.775 11.830 11.792 /
\plot 11.830 11.792 11.898 11.807 /
\plot 11.898 11.807 11.966 11.826 /
\plot 11.966 11.826 12.033 11.843 /
\plot 12.033 11.843 12.112 11.864 /
\plot 12.112 11.864 12.188 11.887 /
\plot 12.188 11.887 12.262 11.910 /
\plot 12.262 11.910 12.338 11.936 /
\plot 12.338 11.936 12.412 11.961 /
\plot 12.412 11.961 12.486 11.989 /
\plot 12.486 11.989 12.560 12.016 /
\plot 12.560 12.016 12.634 12.044 /
\plot 12.634 12.044 12.708 12.073 /
\plot 12.708 12.073 12.783 12.105 /
\plot 12.783 12.105 12.857 12.137 /
\plot 12.857 12.137 12.931 12.169 /
\plot 12.931 12.169 13.005 12.200 /
\plot 13.005 12.200 13.077 12.234 /
\plot 13.077 12.234 13.149 12.268 /
\plot 13.149 12.268 13.221 12.300 /
\plot 13.221 12.300 13.293 12.334 /
\plot 13.293 12.334 13.363 12.370 /
\plot 13.363 12.370 13.432 12.402 /
\plot 13.432 12.402 13.500 12.435 /
\plot 13.500 12.435 13.568 12.469 /
\plot 13.568 12.469 13.633 12.501 /
\plot 13.633 12.501 13.699 12.533 /
\plot 13.699 12.533 13.763 12.565 /
\plot 13.763 12.565 13.824 12.596 /
\plot 13.824 12.596 13.883 12.626 /
\plot 13.883 12.626 13.942 12.653 /
\plot 13.942 12.653 14.002 12.681 /
\plot 14.002 12.681 14.059 12.708 /
\plot 14.059 12.708 14.114 12.734 /
\plot 14.114 12.734 14.169 12.759 /
\plot 14.169 12.759 14.224 12.783 /
\plot 14.224 12.783 14.277 12.806 /
\plot 14.277 12.806 14.330 12.827 /
\plot 14.330 12.827 14.389 12.850 /
\plot 14.389 12.850 14.450 12.874 /
\plot 14.450 12.874 14.512 12.895 /
\plot 14.512 12.895 14.573 12.916 /
\plot 14.573 12.916 14.635 12.935 /
\plot 14.635 12.935 14.700 12.954 /
\plot 14.700 12.954 14.768 12.973 /
\plot 14.768 12.973 14.838 12.990 /
\plot 14.838 12.990 14.910 13.007 /
\plot 14.910 13.007 14.986 13.026 /
\plot 14.986 13.026 15.066 13.043 /
\plot 15.066 13.043 15.151 13.060 /
\plot 15.151 13.060 15.238 13.077 /
\plot 15.238 13.077 15.329 13.094 /
\plot 15.329 13.094 15.422 13.111 /
\plot 15.422 13.111 15.517 13.125 /
\plot 15.517 13.125 15.615 13.142 /
\plot 15.615 13.142 15.708 13.157 /
\plot 15.708 13.157 15.801 13.172 /
\plot 15.801 13.172 15.888 13.187 /
\plot 15.888 13.187 15.970 13.197 /
\plot 15.970 13.197 16.042 13.210 /
\plot 16.042 13.210 16.104 13.219 /
\plot 16.104 13.219 16.154 13.225 /
\plot 16.154 13.225 16.195 13.231 /
\plot 16.195 13.231 16.224 13.236 /
\plot 16.224 13.236 16.241 13.238 /
\plot 16.241 13.238 16.252 13.240 /
\putrule from 16.252 13.240 to 16.256 13.240
\special{ps: grestore}%
%
\linethickness= 0.500pt
\setplotsymbol ({\thinlinefont .})
\special{ps: gsave 0 0 0 setrgbcolor}\putrule from 16.288 13.271 to 16.292 13.271
\plot 16.292 13.271 16.300 13.274 /
\plot 16.300 13.274 16.315 13.276 /
\plot 16.315 13.276 16.341 13.278 /
\plot 16.341 13.278 16.375 13.282 /
\plot 16.375 13.282 16.421 13.288 /
\plot 16.421 13.288 16.478 13.295 /
\plot 16.478 13.295 16.548 13.303 /
\plot 16.548 13.303 16.631 13.314 /
\plot 16.631 13.314 16.722 13.324 /
\plot 16.722 13.324 16.823 13.339 /
\plot 16.823 13.339 16.931 13.352 /
\plot 16.931 13.352 17.046 13.369 /
\plot 17.046 13.369 17.164 13.384 /
\plot 17.164 13.384 17.287 13.401 /
\plot 17.287 13.401 17.410 13.420 /
\plot 17.410 13.420 17.532 13.437 /
\plot 17.532 13.437 17.655 13.456 /
\plot 17.655 13.456 17.776 13.475 /
\plot 17.776 13.475 17.892 13.494 /
\plot 17.892 13.494 18.004 13.513 /
\plot 18.004 13.513 18.112 13.530 /
\plot 18.112 13.530 18.218 13.549 /
\plot 18.218 13.549 18.318 13.568 /
\plot 18.318 13.568 18.415 13.589 /
\plot 18.415 13.589 18.506 13.608 /
\plot 18.506 13.608 18.595 13.627 /
\plot 18.595 13.627 18.677 13.646 /
\plot 18.677 13.646 18.758 13.667 /
\plot 18.758 13.667 18.836 13.688 /
\plot 18.836 13.688 18.910 13.710 /
\plot 18.910 13.710 18.982 13.731 /
\plot 18.982 13.731 19.054 13.754 /
\plot 19.054 13.754 19.122 13.777 /
\plot 19.122 13.777 19.190 13.803 /
\plot 19.190 13.803 19.255 13.828 /
\plot 19.255 13.828 19.321 13.854 /
\plot 19.321 13.854 19.393 13.883 /
\plot 19.393 13.883 19.463 13.917 /
\plot 19.463 13.917 19.535 13.949 /
\plot 19.535 13.949 19.607 13.985 /
\plot 19.607 13.985 19.677 14.021 /
\plot 19.677 14.021 19.748 14.057 /
\plot 19.748 14.057 19.820 14.097 /
\plot 19.820 14.097 19.892 14.135 /
\plot 19.892 14.135 19.964 14.177 /
\plot 19.964 14.177 20.036 14.220 /
\plot 20.036 14.220 20.110 14.264 /
\plot 20.110 14.264 20.185 14.309 /
\plot 20.185 14.309 20.256 14.353 /
\plot 20.256 14.353 20.331 14.400 /
\plot 20.331 14.400 20.405 14.446 /
\plot 20.405 14.446 20.477 14.495 /
\plot 20.477 14.495 20.551 14.541 /
\plot 20.551 14.541 20.623 14.588 /
\plot 20.623 14.588 20.695 14.637 /
\plot 20.695 14.637 20.765 14.683 /
\plot 20.765 14.683 20.834 14.730 /
\plot 20.834 14.730 20.904 14.774 /
\plot 20.904 14.774 20.972 14.819 /
\plot 20.972 14.819 21.040 14.863 /
\plot 21.040 14.863 21.105 14.906 /
\plot 21.105 14.906 21.171 14.948 /
\plot 21.171 14.948 21.234 14.986 /
\plot 21.234 14.986 21.296 15.026 /
\plot 21.296 15.026 21.357 15.062 /
\plot 21.357 15.062 21.419 15.098 /
\plot 21.419 15.098 21.478 15.134 /
\plot 21.478 15.134 21.537 15.166 /
\plot 21.537 15.166 21.594 15.200 /
\plot 21.594 15.200 21.654 15.229 /
\plot 21.654 15.229 21.715 15.261 /
\plot 21.715 15.261 21.778 15.291 /
\plot 21.778 15.291 21.840 15.318 /
\plot 21.840 15.318 21.903 15.346 /
\plot 21.903 15.346 21.969 15.371 /
\plot 21.969 15.371 22.035 15.397 /
\plot 22.035 15.397 22.104 15.422 /
\plot 22.104 15.422 22.176 15.445 /
\plot 22.176 15.445 22.250 15.469 /
\plot 22.250 15.469 22.329 15.492 /
\plot 22.329 15.492 22.409 15.515 /
\plot 22.409 15.515 22.496 15.538 /
\plot 22.496 15.538 22.585 15.560 /
\plot 22.585 15.560 22.676 15.583 /
\plot 22.676 15.583 22.773 15.606 /
\plot 22.773 15.606 22.871 15.627 /
\plot 22.871 15.627 22.972 15.649 /
\plot 22.972 15.649 23.074 15.670 /
\plot 23.074 15.670 23.175 15.691 /
\plot 23.175 15.691 23.273 15.712 /
\plot 23.273 15.712 23.368 15.729 /
\plot 23.368 15.729 23.455 15.746 /
\plot 23.455 15.746 23.535 15.761 /
\plot 23.535 15.761 23.607 15.776 /
\plot 23.607 15.776 23.669 15.786 /
\plot 23.669 15.786 23.717 15.795 /
\plot 23.717 15.795 23.755 15.801 /
\plot 23.755 15.801 23.781 15.805 /
\plot 23.781 15.805 23.800 15.809 /
\plot 23.800 15.809 23.808 15.811 /
\putrule from 23.808 15.811 to 23.812 15.811
\special{ps: grestore}%
%
\linethickness= 0.500pt
\setplotsymbol ({\thinlinefont .})
\special{ps: gsave 0 0 0 setrgbcolor}\plot 23.812 15.811 23.819 15.814 /
\plot 23.819 15.814 23.832 15.816 /
\plot 23.832 15.816 23.855 15.822 /
\plot 23.855 15.822 23.889 15.831 /
\plot 23.889 15.831 23.937 15.843 /
\plot 23.937 15.843 23.999 15.860 /
\plot 23.999 15.860 24.071 15.879 /
\plot 24.071 15.879 24.151 15.900 /
\plot 24.151 15.900 24.238 15.924 /
\plot 24.238 15.924 24.329 15.947 /
\plot 24.329 15.947 24.422 15.972 /
\plot 24.422 15.972 24.513 15.998 /
\plot 24.513 15.998 24.602 16.023 /
\plot 24.602 16.023 24.687 16.046 /
\plot 24.687 16.046 24.767 16.070 /
\plot 24.767 16.070 24.843 16.091 /
\plot 24.843 16.091 24.913 16.112 /
\plot 24.913 16.112 24.981 16.131 /
\plot 24.981 16.131 25.042 16.152 /
\plot 25.042 16.152 25.099 16.171 /
\plot 25.099 16.171 25.154 16.188 /
\plot 25.154 16.188 25.207 16.207 /
\plot 25.207 16.207 25.258 16.224 /
\plot 25.258 16.224 25.305 16.243 /
\plot 25.305 16.243 25.353 16.260 /
\plot 25.353 16.260 25.402 16.281 /
\plot 25.402 16.281 25.451 16.303 /
\plot 25.451 16.303 25.499 16.324 /
\plot 25.499 16.324 25.548 16.347 /
\plot 25.548 16.347 25.599 16.370 /
\plot 25.599 16.370 25.650 16.396 /
\plot 25.650 16.396 25.703 16.423 /
\plot 25.703 16.423 25.756 16.451 /
\plot 25.756 16.451 25.813 16.482 /
\plot 25.813 16.482 25.874 16.516 /
\plot 25.874 16.516 25.936 16.552 /
\plot 25.936 16.552 25.999 16.588 /
\plot 25.999 16.588 26.065 16.626 /
\plot 26.065 16.626 26.130 16.667 /
\plot 26.130 16.667 26.194 16.703 /
\plot 26.194 16.703 26.253 16.739 /
\plot 26.253 16.739 26.306 16.772 /
\plot 26.306 16.772 26.352 16.800 /
\plot 26.352 16.800 26.388 16.823 /
\plot 26.388 16.823 26.416 16.840 /
\plot 26.416 16.840 26.433 16.851 /
\plot 26.433 16.851 26.444 16.857 /
\plot 26.444 16.857 26.448 16.859 /
\special{ps: grestore}%
%
%
\put{\SetFigFont{10}{12.0}{\rmdefault}{\mddefault}{\updefault}$E$} [lB] at 26.194 10.954
%
%
\put{\SetFigFont{10}{12.0}{\rmdefault}{\mddefault}{\updefault}$\varepsilon_0$} [lB] at  8.477  9.5 
%
%
\put{\SetFigFont{10}{12.0}{\rmdefault}{\mddefault}{\updefault}$3\, \varepsilon_0$} [lB] at 15.807  9.5 
%
%
\put{\SetFigFont{10}{12.0}{\rmdefault}{\mddefault}{\updefault}$5\, \varepsilon_0 $} [lB] at 23.422  9.5 
%
%
\put{\SetFigFont{10}{12.0}{\rmdefault}{\mddefault}{\updefault}$N(E)$} [lB] at  5.048 25.718
\put{\SetFigFont{10}{12.0}{\rmdefault}{\mddefault}{\updefault}$\displaystyle \frac{1}{2\pi\ell^2}$} [lB] at  1.810 16.002
%
%
\put{\SetFigFont{10}{12.0}{\rmdefault}{\mddefault}{\updefault}$\displaystyle \frac{2}{2\pi\ell^2}$} [lB] at  1.810 20.860
%
%
%
\linethickness=0pt
\putrectangle corners at  2.642 25.457 and 26.194 10.071
\endpicture}